\documentstyle[12pt]{article}
\begin{document}
\language 1

\centerline{\bf  Special integrals of motion in quantum integrable systems }
\vspace*{1.0cm} 
\centerline{\bf S. Kryukov}
\vspace*{1.0cm}
\centerline {\it Physics Department, University of Lethbridge,}

\centerline {\it 4401 University Drive, Lethbridge, Alberta, T1K 3M4, Canada}

\centerline {\it e-mail: sergei.kryukov@uleth.ca}
\vspace*{0.5cm} 
\centerline{\it Bogoliubov Laboratory of  Theoretical Physics,}
\centerline{\it Joint Institute for Nuclear Research}
\centerline{\it  141980 Dubna (Moscow region),Russia.}
\centerline{\it e-mail: kryukov@itp.ac.ru}
\vspace {10pt}

 {\bf Abstract}  We  investigate quantum integrals of motion
in the  sine-Gordon, Zhiber-Shabat  and  similar systems.
When the  coupling constants in
these models take special values  a new quantum symmetry appears.
In those cases, correlation functions can be obtained, and they have
a power law behavior.
\vspace*{0.25cm}

\noindent Classification codes: PACS 02.10.-v, 03.65.-w \hfill\break

\vspace*{1.0cm} 
\centerline  {\bf 1. Introduction  } 
\vspace*{0.25cm}

Quantization of simple integrable systems is important 
as a  test of quantization methods 
in more general cases.
 Any polynomial field operator must be well defined, and so must be regularized, respecting
the demand of integrability.
In ordinary integrable theory integrals of motion are polynomial field operators, and so have such infinities.
Commutators of those integrals of motion 
also have their own infinities. Our problem  is to determine those operators and  a way to work with polynomial field operators which is free from all
these infinities.
We suppose that know the answer for quantum commutators such integrals of motion. If coupling constants don't take  special values integrals of motion must commute.  So our problem to construct such operators. The  sine-Gordon and Zhiber-Shabat
models are very convenient for investigating this problem.

In [1],   two different methods of quantization are distinguished.
The first is to  construct  the monodromy matrix  ${\cal  T}(\lambda)$ and impose
$$
[{\cal  T}(\lambda),{\cal  T}(\mu)]=0.
$$
The second is to construct the quantum integrals of motion $\hat I_{i}$
and demand
$$
[\hat I_{i},\hat I_{j}]=0.
$$
Classically,   the methods are equivalent. At the quantum level, however, that is not the case.
On the one hand, the  generating function of quantum integrals
of motion $ {\cal  T}(\lambda)$
 is well defined. On the other hand
$\ln {\cal  T}(\lambda)$ is not well defined,
because we have a problem connected with  
divergences [1].  Let us note that for factorization of the $S$ matrix,
we must have quantum integrals of motion (which commute with the Hamiltonian), i.e.
the second method of  quantization is prefereed. This method 
have successfully tasted in quantum integrable sine-Gordon, Zhiber-Shabat system, quantum nonlinear Scrodinger equation, Korteweg-de Vries and modified Korteweg de Vries equations [1].

We have used this second method on the sine-Gordon and Zhiber-Shabat models and have obtained quantum
integrals of motion for arbitrary coupling constants.
If the coupling constants take  special values, an  
interesting situation arises.
Local conservation currents have been found and those
currents have unusual commutation relations, i.e. new algebras.
The generators of the algebras  have two indices. 
This  contrasts with  ordinary
conformal symmetry, where we have the  Virasoro algebra  with one-index generators [2].
Using this new symmetry, we can calculate  correlation functions.

The power law behavior of the resulting correlation functions is  connected with the  violation of the standard equivalence
[3] between the sine-Gordon  and massive Thirring fermionic models. Let us recall basic features of this standard equivalence, and explain probably hidden difficulties of such approach.

We can consider work about equivalence of 
sine-Gordon/Thirring models 
[3] (work of Mandelstam  and in perturbative approach works of Klaiber+Coleman) and explain our position.
In our approach we use only Hamiltonian  quantization. In the works of Mandelstam and Klaiber,  another method has been followed. They solve the normal-ordered Lagrange equations and then check the equal-time  
anticommutators for fermionic fields. Reading those works  carefully reveals an equal-time anticommutator of the non-canonical form
$$
[\hat \psi(x,t),\hat \psi^{+}(y,t)]_{+}=(x-y)^{\sigma}\delta(x-y) \ne \delta(x-y).
$$
We use the Hamiltonian approach and postulate  the  {\it canonical} equal-time anticommutators (for initial operators)  at the beginning of our calculations.

The Hamiltonian approach guarantees the conservation this equal-time anticommutator.
It is easy to see this from the solution of the Hamiltonian equations
$$
\hat \psi(x,t)=\exp \left(-it\hat H \right)~ \hat \psi(x,0)~ \exp \left(it\hat H \right),
$$
where $\hat H$ is the quantum Hamiltonian of the system.
So we  don't have similar problems.
Calculation of integrals of motion demands correct equal-time anticommutators because uses in fact basis property of $\delta$ -  function $\int f(x) \delta(x-y)dx=f(y)$, and so we can not use above anticommutators as in the works of Mandelstam or Klaiber.

In our opinion, the equivalence of these theories
must be investigated
 in the Hamiltonian   approach and a connection between 
integrals of motion for sine-Gordon and massive Thirring models must be found.

\vspace*{1.0cm} 
\centerline  {\bf 2. Two Hamiltonian structures for  the sine-Gordon equation} 
\vspace*{0.25cm}

It is well known that the classical model obeying  the sine-Gordon equation
$$
\partial_{tt}^{2}\varphi(x,t)-
\partial_{xx}^{2}\varphi(x,t)+(m^{2}/\beta)\sin(\beta\varphi(x,t))=0
\eqno(2.1)
$$
is an integrable Hamiltonian system [4].
Let us introduce  equal-time canonical Poisson brackets for
$\varphi(x,t),\pi(x,t)$, where $\pi(x,t)=\partial_{t}\varphi(x,t)$:
$$
\{\varphi(x,t),\pi(y,t)\}=\delta(x-y).\eqno(2.2)
$$
Equation (2.1) can be written in   Hamiltonian form  with the  
Hamiltonian
$$
{\cal H}'=\int_{-\infty}^{+\infty}dx ~~\left(\frac{1}{2}\pi^{2}
+\frac{1}{2}(\partial_{x}\varphi)^{2}
+(m^{2}/\beta^{2})(1-\cos(\beta\varphi))\right).
$$
Integrals of motion  are determined
in terms of  $\varphi(x,t),\pi(x,t)$
by demanding that they commute with ${\cal H}'$, using   the brackets
(2.2). 

The equation of motion also has  a second Hamiltonian structure.
Let us introduce the coordinates $\xi=x+t,~~ \eta=x-t$, and define
$\chi(\xi,\eta)=\varphi((\xi+\eta)/2,(\xi-\eta)/2)$.
In the new coordinates  (2.1) becomes
$$
\partial_{\xi}\partial_{\eta}
\chi(\xi,\eta) -(m^{2}/\beta)\sin(\beta\chi(\xi,\eta))=0.
\eqno(2.3)
$$
The canonical Poisson brackets for $\chi(\xi,\eta)$ are
$$
\{\chi(\xi,\eta),\chi(\xi^{'},\eta)\}=\varepsilon(\xi-\xi^{'}).\eqno(2.4)
$$
Equation (2.3) can therefore be written in   Hamiltonian form with
$$
{\cal H}=(2m^{2}/\beta^{2})\int_{-\infty}^{+\infty}
d\xi(\cos(\beta\chi(\xi,\eta))-1).
 \eqno(2.5)
$$
The Hamiltonian and integrals of motion can be determined at some initial time. The condition on the initial function is a 
holomorphic condition $\partial_{\eta}\chi(\xi,\eta)=0$ ($\eta$ is the time) if we have
$t=i\tau$. We should  determine a quantum field of the theory as
$\hat {\chi}(\xi,\eta)=\exp (-i\eta  \hat {\cal H})~ \hat {\chi}(\xi)~\exp (i\eta \hat {\cal H})$.
Integrals of motion of the Hamiltonian (2.5) and brackets  (2.4)
can be written like
differential polynomials of  $\partial _{\xi}\chi(\xi)$.
We will consider the following boundary condition
$\partial_{\xi} \chi(\xi)$
$(\xi\to \pm\infty)$, and it has some
singularity in a  restricted region of the  real axis $\xi$.
Let us consider $\xi$  as a  complex coordinate and   expand
$\partial_{\xi} \chi(\xi)$  in  some series.
Indeed, any holomorphic function  in the region  $r<|\xi-a|<R$
can be represented by a series in
positive and negative powers of
$(\xi-a)$, the value of $a$-places   where we have a singularity [5].
So we have
$$
\partial_{\xi} \chi(\xi)=\sum_{n\in Z} a_{n} \xi^{-n-1},
$$
and
$$
 \chi(\xi)=q+p\ln\xi -\sum_{n \ne 0}\frac{a_{n}\xi^{-n}}{n}. \eqno(2.6)
$$
We have set $a=0$   without loss of  generality.
Now let us consider quantization of our field.
The commutation relations for the operators are 
$$
[\hat p,\hat q]=1,~~~[\hat  a_{n},\hat  a_{m}]=n\delta_{n+m,0}.
$$
The normal ordering of non-commuting operators obeys
$$
\hat  p \hat  q=:\hat  p \hat q:+1,~~~
\hat  q \hat  p=:\hat  p \hat q:,
$$
$$
\hat  a_{n}\hat  a_{-n}=:\hat  a_{n} \hat  a_{-n}:+n,~~~
\hat  a_{-n}\hat  a_{n}=:\hat  a_{-n} \hat  a_{n}:,~~~n>0.
$$
The relation for field   is the  following
$$
\hat \chi(\xi)\hat \chi(\xi^{'})=: \hat \chi(\xi) \hat \chi(\xi^{'}):
+\ln(\xi-\xi^{'}).
$$
Quantum integrals of motion are differential polynomials of $\hat \chi$
with some coefficients.
These coefficients  should be defined by
$$
[\hat { \cal H},\hat { \cal I}_{k}]=0, \eqno(2.7)
$$
where the quantum Hamiltonian $\hat { \cal H}$ of the theory is
$$
\hat { \cal H}=
\int_{-\infty}^{+\infty}:\exp (\alpha \hat \chi(\xi)):d\xi+
\int_{-\infty}^{+\infty}:\exp - (\alpha \hat \chi(\xi)):d\xi.
$$
When we integrate over $\xi$,  we obtain a singularity.
Indeed, when we expand the  product of ordered operators, we have, for
example
$$
:\exp  (\alpha \hat  \chi(\xi)):~:(\hat \chi(\xi^{'}))^{k}:=
\sum_{n=0}\frac{1}{(\xi-\xi^{'})^{n}}:
\exp (\alpha\hat \chi(\xi))~\hat {P}_{n}^{k}(\xi^{'}):.
$$
We should determine the function at the singularity.
Let us consider the  commutator:
$$
[\hat {\cal  H}, \hat i_{k}(\xi^{'})]=
\hat {\cal H}~ \hat i_{k}(\xi^{'})-\hat i_{k}(\xi^{'})~\hat {\cal H},
$$
where $\hat i_{k}(\xi)$ is the  density of the  integrals of motion.
In the first piece we take the contour of integration
which is  above $\xi'$, and in the second piece we take the contour of
integration which is below $\xi'$.
Then we add integration over $C_{R\to \infty}$    which is equal to zero.
The closed contour can be deformed and we obtain
$$
[\hat {\cal H},\hat i_{k}(\xi^{'})]=\oint_{\xi^{'}}d\xi~
\hat {h}(\xi)~\hat i_{k}(\xi^{'}).
$$
\vspace*{1.2cm} 
\centerline  {\bf 3. Quantum symmetry of integrable theory } 
\vspace*{0.15cm}
Using the  definition of the commutator, we can calculate the integrals of
motion.
If the coupling constants take generic values, we have
$$
\hat i_{4}=
~~:(\partial \hat \chi)^{4}:+\frac{(\alpha^{4}-6\alpha^{2}+4)}{\alpha^{2}}:(\partial^{2}\hat \chi)^{2}:,
$$
$$
\hat i_{6}=
~~:(\partial \hat \chi)^{6}:+\frac{5}{3}
\frac{(-\alpha^{4}+8\alpha^{2}-4)}{\alpha^{2}}
:(\partial {\hat \chi})^{3}\partial^{3}\hat \chi:
+
$$
$$
+\frac{(3\alpha^{8}-40\alpha^{6}+155\alpha^{4}-160\alpha^{2}+48)}{6\alpha^{4}}:(\partial^{3}\hat \chi)^{2}:,
$$
and they satisfy
$$
[\hat {\cal I}_{4},\hat {\cal I}_{6}]=0.
$$
But if the coupling constant takes special values,
some interesting results  appear. We have obtained
$$
[\hat { \cal H},\hat J^{k}(\xi)]=0,
$$
if the coupling constant $\alpha=1 $  in the  sine-Gordon theory.
Here we found
$$
\hat J^{k}(\xi)={{1}\over{(k+1)}}(:(\partial ^{1+k}_{\xi} \exp \hat \chi (\xi))
\partial _{\xi} \exp -\hat \chi (\xi)):
$$
$$
+{{1}\over{(k+2)(k+1)}}:
(\partial ^{2+k}_{\xi} \exp \hat \chi (\xi)) \exp -\hat \chi (\xi)):,~~~k \in {\sf N}.
$$
Similar cases are  listed in the Appendix.

We can calculate the commutation relations between the $\hat J_{n}^{k}$
$$
[\hat J_{n}^{k},\hat J_{m}^{p}]=
(k+1)!\sum_{l'=0}^{k} \frac{  \hat J_{n+m}^{p+l'}}{l'!(k+1-l')!}
\prod_{j'=0}^{ k-l'}(m+p+1-j')-
$$
$$
-(p+1)!\sum_{l=0}^{p} \frac{  \hat J_{n+m}^{k+l}}{l!(p+1-l)!}
\prod_{j=0}^{p-l}(n+k+1-j)
$$
$$
+\delta_{n+m,0}\frac{(p+1)!(k+1)!}{2(k+p+3)!}\times
$$
$$
\times \left((-1)^{k+1}\prod_{j=0}^{2+k+p}(n+k+1-j)- (-1)^{p+1 }
\prod_{j=0}^{2+k+p}(m+p+1-j) \right ),
$$
where
$\hat J^{k}_{n} = \frac{1}{2\pi i}\oint_{0}\hat  J^{k}(\xi)\xi^{k+1+n}d\xi$.
This is a new type of infinite-dimensional algebra, 
with central
extension. Using this quantum symmetry we can calculate  the correlation
functions in  a system of this sort.

\vspace*{1.0cm} 
\centerline  {\bf 4. Evolution of operators in the Heisenberg picture} 
\vspace*{0.25cm}

In the  Heisenberg picture, the  operator $\hat A(\xi,\eta)$
can be obtained from  the operator  $\hat A(\xi)$:
$$
\hat  A(\xi,\eta)=\exp (-i\eta \hat { \cal H})~\hat A(\xi)\exp (i\eta \hat { \cal H}).
$$
The correlation function of  the operators has the form
$$
\langle 0|
\hat A_{1}(\xi_{1},\eta_{1})\cdots
\hat A_{n}(\xi_{n},\eta_{n}) |0
\rangle =
$$
$$
\langle  0|
\exp
(-i\eta_{1} \hat { \cal H})~\hat A_{1}(\xi_{1})\exp(i\eta_{1} \hat { \cal H})\cdots
\exp(-i\eta_{n} \hat { \cal H})~\hat A_{n}(\xi_{n})\exp (i\eta_{n} \hat { \cal H})
|0\rangle ,
$$
where
 $|0\rangle $ is the vacuum of the theory.
If $\hat A(\xi,\eta)$ is a  complicated function of
operator  $\hat \chi$,
we should fix a certain  ordering.
We choose, for example,
$$
\exp (\alpha \hat \chi(\xi,\eta))=
\exp (-i\eta \hat { \cal H}) :\exp (\alpha \hat\chi(\xi)):\exp (i\eta \hat { \cal H}),
$$
$\hat \chi (\xi)$ is the field of theory and it has the decomposition
(2.6).
The normal ordering   : : was  discussed in  section 2.
\vspace*{1.0cm} 
\centerline  { } 
\vspace*{0.25cm}
\centerline{\bf 5. One-point correlation function and properties of the vacuum}

Let us consider the  one-point correlation function
$$
\langle  \hat J(\xi,\eta)\rangle =
\langle 0|\exp (-i\eta \hat{ \cal H})~ \hat J(\xi)\exp (i\eta \hat { \cal H})~|0\rangle =
\langle 0| \hat  J(\xi)|0\rangle.
$$
where $\hat J(\xi,\eta)$-is conservation current in the theory.
In a space-invariant theory, any one-point correlation function is equal
to
a constant $\langle \hat  J(\xi,\eta)\rangle =C$.
This property means that
$$
\langle 0| \hat J_{n}|0\rangle =\delta _{n,0}C,~~~~n=0,\pm1,\pm2...~.
$$
And we have the  property that $\langle 0|$ and $\hat J_{n} |0\rangle$
are orthogonal to each other for all  $ n\ne0$.
We can obtain this property if we assume  that
$$
  \hat J_{n}|0\rangle =0,~~~~n \in \sf N.
$$
In our case we have
$$
  \hat J_{n}^{k}|0\rangle =0,~~~~n \in {\sf N};~~~k \in {\sf N}. \eqno(5.1)
$$
We should have the properties of invariance of vacuum as follows:
$$
\hat { \cal  P}|0\rangle =0, \eqno(5.2)
$$
where 
$ \hat { \cal  P} $-is the operator of translation  ($ \hat { \cal  P}=\frac{1}{2}
\int_{-\infty}^{+\infty}:(\partial_{\xi}\hat \chi(\xi))^2:d\xi $).
We know that
$$
[\hat { \cal  P}, \hat J^{k}_{n}]=(k+n+1) \hat J^{k}_{n-1}. \eqno(5.3)
$$
Now it is easy to obtain
$$
\hat J^{k}_{0}|0\rangle =0,~~
\hat  J^{k}_{-1}|0\rangle =0,~~
\hat  J^{k}_{-2}|0\rangle =0,...~~
\hat  J^{k}_{-k-1}|0\rangle =0,
$$
and for the  left vacuum:
$$
\langle 0|\hat J^{k}_{0}=0,~~
\langle 0|\hat  J^{k}_{1} =0,~~
\langle 0|\hat  J^{k}_{2}=0,...~~
\langle 0|\hat  J^{k}_{k+1} =0.
$$
It is easy to understand that
$$
\langle 0| 
\hat A_{1}(\xi_{1},\eta_{1})....
 \hat A_{n}(\xi_{n},\eta_{n}) \hat J_{p}^{1}|0\rangle =0,~~~ p=0,\pm1,\pm2.
$$
We should know the commutator
$[ \hat J_{n}^{1}, \hat A_{k}(\xi_{k},\eta_{k})]$.
We can calculate this commutator if $\hat A(\xi)=~:\exp (\alpha \hat \chi(\xi))$:
$$
[\hat J^{1}_{n}, :\exp (\alpha \hat \chi (\xi)): ] =
$$
$$
=\xi^{n+2} (\hat J^{1}_{-2},
:\exp \alpha \hat \chi (\xi):)
- \{\alpha \xi^{n+1} (n+2) \partial _{\xi}
+{{(\alpha ^{3}-\alpha) }\over{6}} (n+2) (n+1) \xi^{n} \}
:\exp \alpha \hat \chi (\xi):.
$$

Now we  write  the differential equations for the correlation  function
$$
\sum_{j=1}^{n} \xi_{j}^{p+2} \langle 0|\hat A_{1}(\xi_{1},\eta_{1})...
\exp (-i\eta_{j} \hat { \cal H})~(J_{-2}^{1},
:\exp \alpha_{j} \hat \chi (\xi_{j}):)\exp (i\eta_{j} \hat { \cal H})~...
 \hat A_{n}(\xi_{n},\eta_{n})|0\rangle =
$$
$$
={\cal D}_{p}^{1}f,~~~p=0,\pm1,\pm2,
$$
where ${\cal D}_{p}^{1}$ are known differential operators.
For the two-point correlation function, we have three equations
$$
\alpha(\xi_{2}-\xi_{1})\partial_{\xi_{1}}f+\beta(\xi_{1}-\xi_{2})\partial_{\xi_{2}}f
-\frac{(\alpha^{3}-\alpha+\beta^{3}-\beta)f}{3}=0,
$$
$$
\alpha(\xi_{2}-\xi_{1})(2\xi_{1}+\xi_{2})\partial_{\xi_{1}}f
+\beta(\xi_{1}-\xi_{2})(\xi_{1}+2\xi_{2})\partial_{\xi_{2}}f
$$
$$
-((\alpha^{3}-\alpha)\xi_{1}+(\beta^{3}-\beta)\xi_{2})f=0,
$$
$$
\alpha(\xi_{2}-\xi_{1})(3\xi_{1}^{2}+2\xi_{1}\xi_{2}+\xi_{2}^{2})\partial_{\xi_{1}}f
+\beta(\xi_{1}-\xi_{2})(\xi_{1}^{2}+3\xi_{2}^{2}+2\xi_{1}\xi_{2})\partial_{\xi_{2}}f
$$
$$
-(2(\alpha^{3}-\alpha)\xi_{1}^{2}+2(\beta^{3}-\beta)\xi_{2}^{2})f=0.
$$
They have a solution (if we set $\beta=-\alpha$)
$$
f(\xi_{1},\xi_{2},\eta_{1},\eta_{2})=C(\alpha)
G(\eta_{1},\eta_{2})(\xi_{1}-\xi_{2})^{(1-\alpha^{2})},
$$
where $G(\eta_{1},\eta_{2}),C(\alpha)$ are arbitrary functions.
The equation (2.3) in light-cone coordinates has  the symmetry
$(\xi \leftrightarrow \eta)$,
and so we have
$f(r) \sim r^{2(1-\alpha^{2})} $
where $r$ is the distance between the points.
We can investigate other Ward identities from another current
$\hat J^{2}(\xi)$ and obtain the same formula for $f(r)$.

Similar conservation currents  exist  in other models (see  the Appendix).
Now let us consider integrals of motion in systems with Hamiltonian
$$
\hat { \cal H}_{k}=
\int_{-\infty}^{+\infty}:\exp(\alpha \hat \chi(\xi)):d\xi+
\int_{-\infty}^{+\infty}:\exp (k \alpha \hat \chi(\xi)):d\xi.
$$
We have obtained  nontrivial integrals of motion if
$k=-1,-2,
\frac{1}{\alpha^{2}},
\frac{2}{\alpha^{2}},
\frac{4}{\alpha^{2}}$ 
and local  conservation currents (see Appendix).
When $k=-1$
  we have the sine-Gordon theory and if $k=-2$, the Zhiber-Shabat theory.
The theories with other $k$  have no classical limit 
($\alpha \to 0$).

\vspace*{1.0cm} 
\centerline  {\bf 6. Conclusion } 
\vspace*{0.25cm}

Computer calculations confirm an absolutely general behavior of quantum
integrable systems with exponential interactions.
We have obtained some   examples of local conservated currents (densities) in all known models.
Correlation functions in these cases can  probably  be obtained by a similar 
procedure. We have obtained a new infinite-dimensional symmetry in quantum
integrable theories, and 
we  calculated two-point correlation function
using this symmetry.

There is an interesting connection between this calculation and a 
similar calculation in fermionic models [6],
where we used the Hamiltonian method 
to solve the  massless Thirring model and obtained an
anomaly for conservation of the energy-momentum tensor. We in fact reveal the 
hidden problem in fermionic anticommutators by demonstrating on obvious 
problem-- the anomaly. So the massless Thirring model is not a  conformal field theory (solvable theory) and we can not  use Coleman's way of proving  the equivalence of the massive Thirring model and sine-Gordon theories using perturbative methods.

\vspace*{1.0cm} 
\centerline  {\bf  Acknowledgements } 
\vspace*{0.25cm}

 This work was supported by a NATO Science
Fellowship, the University of Lethbridge, and by NSERC of Canada.
We thank  Mark Walton and Jian-Ge Zhou for discussions.

\vspace*{1.0cm} 
\centerline  {\bf Appendix A.}
\vspace*{0.25cm}

Here we report the  arrangement and quantities $m_{n}$ of 
nontrivial currents at  levels $n$.
A current which is a differential of another current is trivial.
The results of  computer calculations are listed  in the Table below.
We investigate the space of quantum integrals of motion for the Hamiltonian $\hat {\cal H}_{k}$.
This problem can move to investigation of system 
$S_{n}$ ordinary linear equations with parameters $\alpha, k$, degeneration of this system can  correspond to appear the current at level $n$.
Determinant of the system for $n=4, 6$ have a form
$$
\det S_{4}=(\alpha^{2}-1)(1-k)^{2}(\alpha^{2} k+2)^{2}(k+1)(-\alpha^{2}k+2),
$$
$$
\det S_{6}=
$$
$$
(\alpha-1)^{2}(\alpha+1)^{2}(3\alpha^{2}-2)
(\alpha^{2}-3)(1-k)^{4}(k\alpha^{2}+2)^{4}(k+1)(k+2)\times
$$
$$
\times (2k+1)(-k\alpha^{2}+2)(-k\alpha^{2}+4)(-k\alpha^{2}+1)
(-2\alpha^{2}k^{2}+k\alpha^{2}-2k-3)
$$
After we find solutions for current when 
$\det S_{n}=\det S_{4}=0$,
which demand some conditions for parameters $\alpha,k$.

\vskip 10 pt
\centerline{ Table}
$$
\begin {tabular}{|l|l|l|l|l|l|l|l|l|l|l|l|}
\hline
$\alpha$ & k&  $m_{2}$ & $m_{3}$ & $m_{4}$ & $m_{5}$ &$m_{6}$ & $m_{7}$ & $m_{8}$ & $m_{9}$ & $m_{10}$&$m_{n}$-conjecture\\
\hline
1& -1   & 0  &1   & 1  & 1  & 1  & 1  & 1  & 1  & 1&
$m_{n}=1$, for $n>2$               \\
\hline
$\sqrt 2 $& -1   & 1  & 0  & 1  & 0  & 1  & 0  & 1  &  0  &1 &
$m_{2p}=1$,  $p \in \sf N$ \\
\hline
$\sqrt 3$& -2   & 0  & 0  &0   &0   & 1  & 0  & 0  & 0  & 1 &
$m_{6+4(p-1)}$, $p \in \sf N$  \\
\hline
$\sqrt {\frac{2}{3}}$ & -2 & 0 & 0 & 0 & 0 &1  &0  & 1 & 0&1&
$m_{6+2(p-1)}$, $p \in \sf N$      \\
\hline
$\sqrt 3 $&$\frac{2}{\alpha^{2}}$   & 0 & 0 & 0 & 0 & 0 & 0 & 1 & 0 & 1&
 $m_{8+2(p-1)}$, $p \in \sf N$      \\
\hline
$\sqrt 3 $&$\frac{1}{\alpha^{2}}$   & 0 & 0 & 0 & 0 & 0 & 0 &  1&  0 &1& 
 $m_{6+4(p-1)}$, $p \in \sf N$       \\
\hline
$\sqrt{\frac{2}{3}} $& $\frac{1}{\alpha^{2}}$&  0&   0& 0 & 0 &  1&  0&  1&  0& 1& 
$m_{6+2(p-1)}$,  $p \in \sf N$    \\
\hline
\end {tabular}
$$

\vspace*{1.0cm} 
\centerline  {\bf References } 
\vspace*{0.25cm}

[1] R. Sasaki, M. Omote, M. Sakagami, I. Yamanaka, Phys. Rev. D, 1987,

~~v.35, n.8, p.2423.

~~R. Sasaki, I. Yamanaka, Comm. Math. Phys., 1987, v.108, p.691.

~~R. Sasaki, I. Yamanaka, Adv. Stud. in Pure. Math., v.16, 1988, p.271.

[2] A. A. Belavin, A. M. Polyakov, A. B. Zamolodchikov, 1984, Nucl. Phys, B241, 333.

[3] B. Klaiber, in Lectures in Theoretical Physics 

Lectures
delivered at the Summer Institute for Theoretical 
Physics

University of Colorado, Boulder, 1967, ed A.Barut, W.
Brittin, Gordon and Breach, New York, 1968, v.X, part A, p. 141-176.

S. Coleman, Phys. Rev. D11 (1975) 2088.

S. Mandelstam, Phys. Rev. D11 (1975) 3026.

[4] L. A. Takhtadzhan, L. D. Faddeev, Hamilton approach in soliton theory. Moscow.,1986.

[5] B. V. Shabat, Introduction to complex analysis. Moscow.,~~ 1976.

[6] S. Kryukov, Dirac quantization of massless Thirring model: energy momentum tensor anomaly, Preprint.(nonlin-SI-0403016)

\end{document}